\documentclass[a4paper]{jpconf}
\usepackage{amsmath}
\usepackage{amssymb}
\usepackage{hyperref}
\usepackage{color}
\usepackage{graphicx}

\begin{document}
\title{Secular Outflows from Long-Lived Neutron Star Merger Remnants}
\author[0000-0001-6982-1008]{David Radice}
\address{Institute for Gravitation \& the Cosmos, The Pennsylvania State University, University Park, PA 16802}
\address{Department of Physics, The Pennsylvania State University, University Park, PA 16802}
\address{Department of Astronomy \& Astrophysics, The Pennsylvania State University,University Park, PA 16802}
\ead{david.radice@psu.edu}

\author[0000-0002-2334-0935]{Sebastiano Bernuzzi}
\address{Theoretisch-Physikalisches Institut, Friedrich-Schiller-Universit{\"a}t Jena, 07743, Jena, Germany}

\begin{abstract}
  We study mass ejection from a binary neutron star merger producing
  a long-lived massive neutron star remnant with general-relativistic
  neutrino-radiation hydrodynamics simulations. In addition to outflows
  generated by shocks and tidal torques during and shortly after the
  merger, we observe the appearance of a wind driven by spiral density
  waves in the disk.  This spiral-wave-driven outflow is predominantly
  located close to the disk orbital plane and have a broad distribution
  of electron fractions. At higher latitudes, a high electron-fraction
  wind is driven by neutrino radiation. The combined nucleosynthesis
  yields from all the ejecta components is in good agreement with Solar
  abundance measurements.
\end{abstract}

% ============================================================================
\section*{Introduction}
% ============================================================================
Neutron star merger outflows are thought to be one of the main
astrophysical sites of production of heavy r-process elements
\cite{Thielemann:2017acv}. Strong evidence linking mergers and r-process
nucleosynthesis is provided by AT2017gfo, the kilonova associated with
the neutron star merger GW170817. In particular, models of the
light-curve and spectra of AT2017gfo require the presence of rare earth
elements in the ejecta \cite{Coulter:2017wya, Kasliwal:2017ngb,
Chornock:2017sdf, Villar:2017wcc, Perego:2017wtu, Kawaguchi:2018ptg}.
More recently, kilonova-like signals were observed following the long
gamma-ray bursts 211211A \cite{Rastinejad:2022zbg, Troja:2022yya,
Yang:2022qmy} and 230307A \cite{Sun:2023rbr, JWST:2023jqa,
Yang:2023mqt}, which were also interpreted as binary neutron star
mergers \cite{Gottlieb:2023sja}. However, with the possible exception of
strontium \cite{Watson:2019xjv, Perego:2020evn} and tellurium
\cite{Hotokezaka:2023aiq}, it has not been possible to identify specific
heavy r-process elements in AT2017gfo, or other kilonovae. As such,
nucleosynthesis yields need to be estimated from merger simulations.

A major difficulty for theoretical models is that the outflows in
neutron star mergers are generated through a variety of mechanisms
operating over different timescales \cite{Shibata:2019wef}. Some of the
ejecta is generated by tidal torques and shocks over a timescale of
milliseconds during and shortly after the merger
\cite{Hotokezaka:2012ze, Korobkin:2012uy, Bauswein:2013yna,
Sekiguchi:2015dma, Radice:2016dwd, Lehner:2016lxy, Dietrich:2016hky,
Radice:2018pdn, Vincent:2019kor, Nedora:2020hxc, Foucart:2020qjb,
Combi:2022nhg, Schianchi:2023uky} the so-called dynamical ejecta.  Winds
from the remnant operate on a timescale of seconds, the so-called
secular ejecta \cite{Perego:2014fma, Martin:2015hxa, Fernandez:2013tya,
Fujibayashi:2017puw, Siegel:2017jug, Fernandez:2018kax, Miller:2019dpt,
Most:2021ytn, Curtis:2022hjy, Fahlman:2022jkh, Sprouse:2023cdm,
Just:2023wtj, Combi:2023yav}. Most studies have focused on either the
secular, or the dynamical ejecta. A notable exception are the recent
works by Kiuchi and collaborators \cite{Kiuchi:2022nin}, considering the
dynamical and secular ejecta from a binary neutron star system producing
a black hole shortly after merger, and of Hayashi et
al.~\cite{Hayashi:2021oxy}, considering the long-term evolution of a
neutron-star black hole merger remnant. However, such simulations are
extremely challenging because, while fully general-relativistic codes
are required to model the merger, the large computational costs of these
simulations make them unsuitable for long-term integration. On the other
hand, various groups have successfully modeled the secular ejecta,
particularly from black-hole torus systems, while neglecting the
back-reaction of the metric, or in Newtonian gravity. This approach is
viable for remnants forming black-holes, but not for mergers producing
long-lived massive neutron star remnants. In the latter case, a major
outflow component is the spiral-wave wind, which is driven by
hydrodynamic instabilities inside the remnant. Capturing such
instabilities requires the same-kind of full-3D simulations with
dynamical spacetime that are used to model the merger phase.
Understanding such mergers is important, since it is believed that a
substantial fraction of mergers produce long-lived remnants. GW170817 is
also likely to have resulted in the formation of a meta-stable remnant
\cite{Rezzolla:2017aly, Li:2018hzy, Metzger:2018qfl, Ai:2018jtv,
Nedora:2019jhl, Ciolfi:2020wfx, Mosta:2020hlh, Combi:2023yav,
Curtis:2023zfo, Kawaguchi:2023zln}.

In Ref.~\cite{Radice:2023zlw}, we have reported on the first ab-initio,
general-relativistic neutrino-radiation hydrodynamics simulations of a
long-lived neutron star merger remnant to secular timescales. We
reported on the structure of the remnant and of the disk and on their
evolution in the first 100~ms after the merger. Here, we study the mass
ejection and nucleosynthesis yield from our models. We find that a
massive outflow is launched by the spiral-wave wind mechanism, in
agreement with previous studies with more approximate neutrino
treatment. We discuss the geometry and composition of the outflows.
Finally, we show that our model predicts an r-process nucleosynthesis
pattern in good agreement with Solar abundance measurements.

\begin{figure}
  \centering
  \includegraphics[scale=0.9]{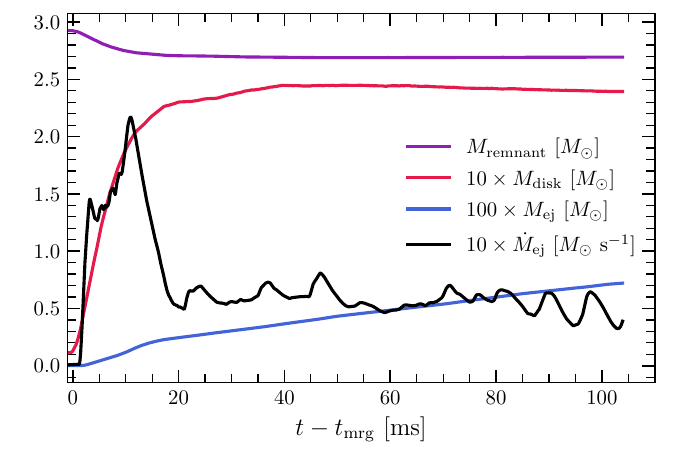}
  \caption{Baryonic rest mass of the remnant ($\rho > 10^{13}\ {\rm g}\
  {\rm cm}^{-3}$), of the disk ($\rho < 10^{13}\ {\rm g}\ {\rm
  cm}^{-3}$), and of the ejecta ($-h u_t \geq h_{\min}$), and mass
  outflow rate as a function of time from merger. The disk is initially
  formed of material expelled from the remnant on a timescale of
  ${\sim}20\ {\rm ms}$. It then evolves on a secular timescale under the
  effect of the spiral-wave instability. The disk mass evolution curve
  has been smoothed using a square window with width of $1\ {\rm ms}$.}
  \label{fig:remnant}
\end{figure}

% ============================================================================
\section*{Methods}
% ============================================================================
We consider a binary with component masses $1.35\ M_\odot$ and $1.35\
M_\odot$. We adopt the HS(DD2) equation of state \cite{Typel:2009sy,
Hempel:2009mc}, which predicts a maximum nonrotating neutron-star mass
of $2.42\ M_\odot$ and the radius of a nonrotating $1.4\ M_\odot$
neutron star to be $R_{1.4} = 13.2\ {\rm km}$. We construct irrotational
initial data with the \texttt{Lorene} pseudo-spectral code
\cite{Gourgoulhon:2000nn}.  The initial separation is of 50~km.
Evolutions are carried out using the gray moment-based (M1)
general-relativistic neutrino-radiation hydrodynamics code
\texttt{THC\_M1} \cite{Radice:2012cu, Radice:2013hxh, Radice:2013xpa,
Radice:2015nva, Radice:2021jtw}, which is based on the \texttt{Einstein
Toolkit} \cite{Loffler:2011ay, EinsteinToolkit:2023_05}. For the
simulations discussed here, we use the \texttt{Carpet} adaptive
mesh-refinement (AMR) driver \cite{Schnetter:2003rb, Reisswig:2012nc},
which implements the Berger-Oilger scheme with refluxing
\cite{Berger:1984zza, 1989JCoPh..82...64B}, and evolve the spacetime
geometry using the \texttt{CTGamma} code \cite{Pollney:2009yz,
Reisswig:2013sqa}, which solves the Z4c formulation of Einstein's
equations \cite{Bernuzzi:2009ex, Hilditch:2012fp}. The simulations
reported here neglect magnetic fields. On the one hand, it is known that
if the stars are originally endowed with magnetar-level fields, the
magnetic stresses can qualitatively affect the outflow dynamics
\cite{Combi:2023yav}. On the other hand, simulations starting with
realistic initial fields and employing a sophisticated treatment to
capture unresolved turbulence and dynamo action show that the magnetic
fields are unlikely to be dynamically significant over the timescales
considered here \cite{Palenzuela:2022kqk, Aguilera-Miret:2023qih}. In
particular, the spiral-wave wind is expected to dominate the angular
momentum transport in the disk, for the values of turbulent viscosity
inferred by those simulations \cite{Nedora:2019jhl}. The results of
additional simulations, including with more equations of state and with
the inclusion of angular momentum transport due to turbulent viscosity
will be reported elsewhere.

The evolution grid employs 7 levels of AMR. Our simulation is performed
at two resolutions with the finest grid having finest grid spacing of $h
= 0.167\ G M_\odot/c^2 \simeq 247\ {\rm m}$ and $h = 0.125\ G
M_\odot/c^2 \simeq 185\ {\rm m}$, respectively denoted as LR and SR
setups. Here, we report
the results from the higher resolution simulations. There
lower-resolution data is in good quantitative and qualitative agreement.

\begin{figure}
  \centering
  \includegraphics[trim={0 0 0 0.65cm},clip,scale=0.8]{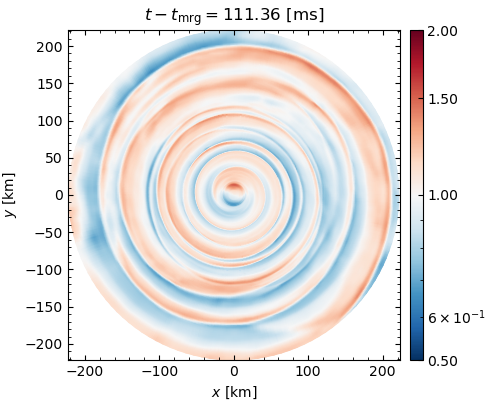}
  \caption{Density contrast in the disk $111.36\ {\rm ms}$ after merger.
  The spiral-wave in the disk is generated by the $m=2$ deformation of
  the remnant (the so-called bar mode) and persists throughout our
  simulation. }
  \label{fig:spiral}
\end{figure}

% ============================================================================
\section*{Results}
% ============================================================================

The binary considered in this study results in the formation of a
remnant that is dynamically stable, as documented in,
e.g.~\cite{Radice:2018xqa}.  Figure~\ref{fig:remnant} shows the
evolution of the baryon mass density inside the remnant massive neutron
star and in the disk. Following a standard convention, we flag material
as belonging to the remnant massive neutron star if its density is
larger than $10^{13}\ {\rm g}\ {\rm cm}^{-3}$. Material with lower
density and situated within a cubical box of half-diameter $160 \ G
M_\odot/c^2 \simeq 236\ {\rm km}$ is consider as belonging to the disk.
The latter is formed of matter that is squeezed out of the collisional
interface between the neutron stars during and shortly after merger
\cite{Radice:2018pdn, Zenati:2023lwh}. After ${\sim}20{-}30\ {\rm ms}$
mass ejection from the central remnant terminates. The subsequent disk
mass evolution is primarily driven by mass loss to outflows with
negligible accretion onto the central object.

\begin{figure}
  \centering
  \includegraphics[scale=0.9]{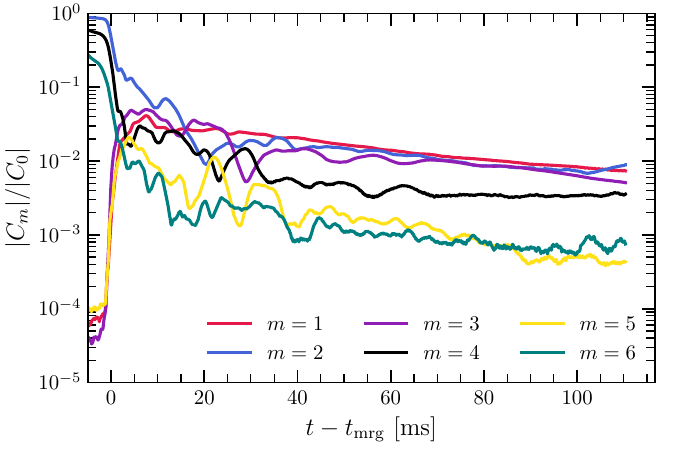}
  \caption{Normalized density modes on the $xy$-plane (see Eq.
  \ref{eq:modes}). The $m=1,2,3$ density modes are the dominant modes.
  They are driven by the low-$T/|W|$ instability after merger. The
  curves have been smoothed using a square windows with width of $1\ {\rm ms}$.}
  \label{fig:modes}
\end{figure}

The ejecta are extracted on a sphere of coordinate radius $R = 300\ G
M_\odot/c^2 \simeq 443\ {\rm km}$, only material with outgoing radial
velocity and that is unbound according to the Bernoulli
condition\footnote{That is with $-h\, u_t \geq h_{\rm min}$, where $h$
is the specific enthalpy, $h_{\min}$ is the minimum specific enthalpy of
the equation of state, and $u$ is the fluid four-velocity
\cite{Foucart:2021ikp}.} is considered in our analysis. Our simulations
show two distinct mass ejection components: a first pulse of material
(${\sim}3\times 10^{-3}\ M_\odot$) component ejected during and shortly
after the merger, followed by a steady wind with $\dot{M}\simeq 0.05\
M_\odot\ {\rm s}^{-1}$. The first ejecta component comprises the tidal
tails of the stars and the shocked ejecta launched shortly after merger
\cite{Radice:2018pdn}. The second component is dominated by the
spiral-wave wind \cite{Nedora:2019jhl}. However, we also observe the
emergence of a neutrino-driven wind at high latitudes
\cite{Nedora:2020hxc, Radice:2021jtw}.

The spiral-wave wind dominates the overall mass outflow rate in the
postmerger. This wind is the result of hydrodynamic and gravitational
torques exerted by the remnant on the disk. The former is deformed as a
result of the so-called low-$T/|W|$ instability \cite{Rampp:1997em,
Ou:2006yd, Corvino:2010yj, Paschalidis:2015mla, Radice:2016gym,
LongoMicchi:2023khv}.  To diagnose this we show in Fig.~\ref{fig:spiral}
the density contrast on the orbital plane of the disk:
\begin{equation}
  \frac{\rho - \langle \rho \rangle}{\langle \rho \rangle},
\end{equation}
where $\rho$ is the fluid rest-mass density and $\langle\cdot\rangle$
denotes average in the azimuthal direction. The $m=2$ pattern in the
remnant is clearly visible at the center, as are the spiral waves. These
waves carry angular momentum outwards and drive the mass outflow, as
anticipated by previous long-term post-merger simulations with
simplified neutrino physics \cite{Nedora:2019jhl, Nedora:2020hxc}. This
spiral wave pattern forms shortly after merger and persists for the
entire duration of our simulations. The spiral-wave wind is
predominantly emitted close to the disk orbital plane and has a broad
distribution of electron fractions (see Fig.~\ref{fig:hist} below).

\begin{figure}
  \centering
  \includegraphics[scale=0.9]{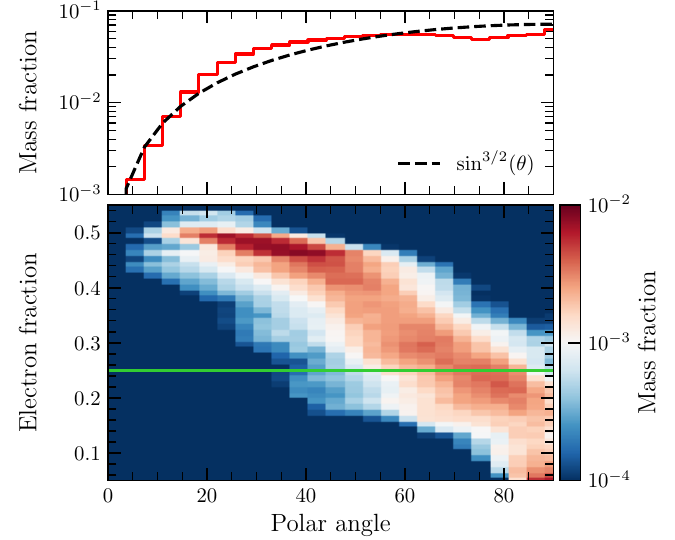}
  \caption{Angular distribution and electron fraction of the ejecta.
  Most of the ejecta is equatorial and has a broad range of electron
  fractions. The polar ejecta is less massive and characterized by
  significantly larger electron fraction $Y_e \gtrsim 0.4$.}
  \label{fig:hist}
\end{figure}

Figure~\ref{fig:modes} shows the evolution of the density modes on the
orbital plane,
\begin{equation}\label{eq:modes}
  C_m = \int_{z=0} e^{-{\rm i}\, m\, \varphi} \rho\, W\, \sqrt{\gamma}\,
  {\rm d}^2 x,
\end{equation}
as a function of time. The odd-$m$ modes are initially zero, because
equal mass binaries have a $\pi$-symmetry\footnote{Meaning that they are
symmetric under a rotation of $180^\circ$ about the $z$-axis.}.
However, the symmetry is broken by turbulence during merger and
subsequently the odd-$m$ modes are excited by the low-$T/|W|$
instability \cite{Radice:2016gym, Espino:2023llj}.  The even $m$ modes
are present from the beginning, particularly the $m=2$ mode. In absence
of the low-$T/|W|$ instability these modes would be efficiently damped
by gravitational radiation \cite{Bernuzzi:2015opx}. However, in our
simulation they are continuously driven by the instability and saturate
at a roughly constant amplitude or decay slowly.  The one-arm ($m=1$
mode) and the triaxial ($m=3$ mode) instabilities dominate the remnant
dynamics from ${\sim 20}$~ms to ${\sim}60$~ms postmerger. Afterwards,
the $m=1,2,3$ modes equally contribute to the remnant dynamics.  This
has been observed in both isolated neutron star \cite{Ou:2006yd,
Manca:2007ca} and binary mergers \cite{Radice:2016gym} simulations,
although the details of the evolution are dependent on the equation of
state, the compactness and possibly the microphysics
\cite{Espino:2023llj}.  The gravitational waves associated to these
nonaxisymmetric modes carry the imprinting of the equation of state but,
even under optimistic assumption, an observation will likely require
third-generation detectors and a source at distance of ${\sim}30$~Mpc
\cite{Radice:2016gym}.  We caution the reader that, because our code
uses a Cartesian grid, the $m=4$ mode might be artificially enhanced.

While the structure of the wind close to the equator is similar to that
observed in simulations with simplified M0 neutrino transport we have
published in the past \cite{Nedora:2019jhl, Nedora:2020hxc,
Zappa:2022rpd}, the wind at high latitude is substantially enhanced in
our new M1 simulations, as we have also documented for a lower
resolution simulation in \cite{Radice:2021jtw}. The angular structure of
the ejecta is shown in Fig.~\ref{fig:hist}, upper panel.  This figure
should be contrasted with Fig.~1 of \cite{Perego:2017wtu}, where the
same diagnostics is shown from our older simulations using M0 neutrino
transport. In contrast to the M0 simulations, which had angular profiles
well fitted by $\sin^2(\theta)$ -- $\theta$ being the angle from the
equatorial plane -- the new M1 simulations show a shallower profile,
closer to $\sin^{3/2}(\theta)$. The composition of the ejecta is also
quite different: the M0 simulations predict ejecta with electron
fraction $Y_e < 0.4$ at all latitudes, while the newer M1 simulations
predict a broader distribution with $Y_e \lesssim 0.5$. These
differences arise from the fact that the M0 scheme artificially
suppresses neutrino heating in the polar region of the remnant,
suppressing the wind \cite{Radice:2021jtw}.

\begin{figure}
  \centering
  \includegraphics[scale=0.9]{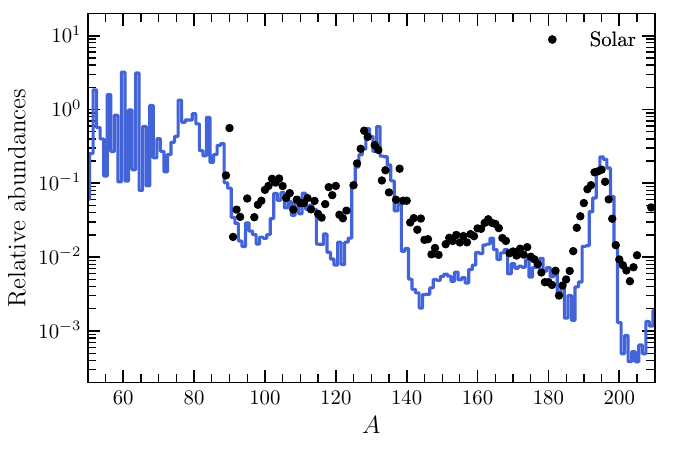}
  \caption{Nucleosynthesis yields of the ejecta (solid line) and Solar
  r-process abundances from Ref.~\cite{Arlandini:1999an} (points). The
  spiral-wave wind produces a full r-process yield.}
  \label{fig:yields}
\end{figure}

Figure~\ref{fig:yields} shows the combined nucleosynthesis yields for
all ejecta components. These have been obtained using the
\texttt{SkyNet} code \cite{Lippuner:2017tyn} with the procedure
discussed in Ref.~\cite{Radice:2018pdn}. We find that our model well
reproduces the Solar r-process pattern. The only significant discrepancy
is in the region with $A\sim 140$, however those can be ascribed to the
particular nuclear-mass model used in our nucleosynthesis calculations
\cite{Vassh:2022nij}.

% ============================================================================
\section*{Conclusions}
% ============================================================================
We have performed long-term simulations of a binary neutron star merger
producing a long-lived remnant with a general-relativistic
neutrino-radiation hydrodynamics code. We find that such events drive
massive outflows through a combination of tidal torques and shocks,
during and shortly after the merger, and spiral-wave and neutrino-driven
winds after the merger. The dynamics of the spiral-wave wind is in good
agreement with predictions from simulations with lower-fidelity neutrino
transport. However, there are substantial differences in the dynamics
and composition of the neutrino-driven wind. In particular, our previous
studies underestimated both the mass and the electron fraction of this
component of the ejecta. The combined nucleosynthesis yield of all
ejecta components is in good agreement with the Solar r-process
abundance pattern, indicating that neutron star mergers producing
long-lived remnants provide a natural environment for the formation of
these elements.

% ============================================================================
\section*{Acknowledgments}
% ============================================================================
DR acknowledges funding from the U.S. Department of Energy, Office of
Science, Division of Nuclear Physics under Award Number(s) DE-SC0021177
and from the National Science Foundation under Grants No. PHY-2011725,
PHY-2020275, PHY-2116686, and AST-2108467.
This research used resources of the National Energy Research Scientific
Computing Center, a DOE Office of Science User Facility supported by the
Office of Science of the U.S.~Department of Energy under Contract
No.~DE-AC02-05CH11231.
SB knowledges funding from the EU Horizon under ERC Consolidator Grant,
no. InspiReM-101043372 and from the Deutsche Forschungsgemeinschaft, DFG,
project MEMI number BE 6301/2-1.
The authors acknowledge the Gauss Centre for Supercomputing
e.V. (\url{www.gauss-centre.eu}) for funding this project by providing
computing time on the GCS Supercomputer SuperMUC-NG at LRZ
(allocation {\tt pn36ge} and {\tt pn36jo}).

% ============================================================================
\section*{References}
% ============================================================================
\bibliographystyle{iopart-num.bst}
\providecommand{\newblock}{}

\end{document}